# Poromechanical behaviour of hardened cement paste under isotropic loading

Siavash Ghabezloo[1,*], Jean Sulem[1], Sylvine Guédon[2],

François Martineau[2], Jérémie Saint-Marc[3]

[1] *Université Paris-Est, UR Navier, CERMES, Ecole Nationale des Ponts et Chaussées, Marne la Vallée, France*

[2] *Université Paris-Est, LCPC, MSRGI, Paris, France*

[3] *TOTAL, Management of Residual Gases Project, Pau, France*

## Abstract

The poromechanical behaviour of hardened cement paste under isotropic loading is studied on the basis of an experimental testing program of drained, undrained and unjacketed compression tests. The macroscopic behaviour of the material is described in the framework of the mechanics of porous media. The poroelastic parameters of the material are determined and the effect of stress and pore pressure on them is evaluated. Appropriate effective stress laws which control the evolution of total volume, pore volume, solid volume, porosity and drained bulk modulus are discussed. A phenomenon of degradation of elastic properties is observed in the test results. The microscopic observations showed that this degradation is caused by the microcracking of the material under isotropic loading. The good compatibility and the consistency of the obtained poromechanical parameters demonstrate that the behaviour of the hardened cement paste can be indeed described within the framework of the theory of porous media.

[*] Corresponding Author : CERMES, Ecole Nationale des Ponts et Chaussées, 6-8 avenue Blaise Pascal, Cité Descartes, 77455 Champs-sur-Marne, Marne la Vallée cedex 2, France
Email: ghabezlo@cermes.enpc.fr

# 1 Introduction

In oil wells, a cement sheath is placed between the rock and the casing for support and sealing purpose. The cement lining is submitted to various thermal and mechanical loadings during the life of the well from the drilling phase to the production phase and finally in the abandon phase when the well must seal the subsurface from the surface, as for instance for storage and sequestration of greenhouse gas. In due course of these solicitations, the cement can be damaged and the mechanical and transport properties can be degraded, this degradation being detrimental to its functions. The knowledge of poromechanical behaviour and permeability of this cement in oil-wells conditions, i.e. under high stress and elevated temperature, is essential for the prediction of the well performance during the exploitation and also prediction of the sealing performance of the well when $CO_2$ storage and sequestration is planned.

Whether the mechanical behaviour of hardened cement paste can be described within the framework of the mechanics of porous media is an open question. The main point of discussion about the possibility to actually consider the cement paste as a poromechanical material is related to the complex microstructure of the material: the porosity of this material manifests itself at different scales so that a part of the water phase can not be considered as a bulk water phase, but as structural water [1]. Recent progress of advanced micromechanical testing methods, such as nano-indentation tests, have provided direct measurements of the elastic properties of the different phases of the microstructure of the hardened cement paste [2][3][4][5][6][7] and has made it possible to estimate the poromechanical properties of this material using the homogenization techniques [6][8][9]. These evaluations of the poromechanical properties are mostly supported by the experimental results obtained using non-destructive methods such as, among others, elastic resonance and ultrasonic wave velocity measurements, but have not been validated against mechanical loading tests results. Indeed very few experimental results of classical poromechanics tests, such as drained, undrained and unjacketed compression tests, can be found in the literature for the hardened cement paste. Some tests have been presented by Heukamp [10] and Heukamp *et al*. [11][12], but these studies are mostly focussed on calcium-leached cement pastes and the effect of chemical degradation on the mechanical properties of hardened cement paste. An important difficulty with the hardened cement paste is its very low permeability which makes the poromechanical tests long and expensive.

In this paper, the question whether the hardened cement paste can be considered as a poromechanical material or not is addressed on the basis of an experimental approach. The mechanical behaviour of the hardened cement paste under isotropic loading is studied through drained, undrained and unjacketed compression tests, where the classical testing procedures have been adapted for low permeability materials. In doing this, the macroscopic behaviour of the material is studied and described with the tools of the mechanics of porous media as it is done classically for saturated porous rocks. The compatibility and the consistency of the experimental results and of the obtained poromechanical parameters are discussed to state whether the behaviour of the hardened cement paste can be described within the framework of the theory of porous media.

The second part of the paper is dedicated to the theoretical aspects of modelling the elastic volumetric behaviour of a porous material which can be heterogeneous and anisotropic at the micro-scale. The general concept of effective stress is presented and discussed and the expressions of the effective stress coefficients for different physical properties of a porous material are derived. A discussion is also presented on the cement paste porosity that should be used in poromechanical formulations. The set of parameters introduced in this section is evaluated in the experimental study. The testing program is presented in the third part of the paper and the obtained results are discussed. Observations of the microstructure of the samples have been performed before and after the mechanical loading tests using an environmental scanning electron microscope (ESEM). These microphotographs demonstrate the evolution of the microstructure during loading and the degradation phenomenon observed in the test results. These microscopic observations and the discussion of the degradation mechanisms are presented respectively in the fourth and fifth parts of the paper.

## 2   Poromechanical background

We present here the framework used to describe the elastic volumetric behaviour of a porous material which can be heterogeneous and anisotropic at the micro-scale. The theoretical basis of the formulation has been presented in many earlier studies. Among them, one can refer to the milestone papers and textbooks of Biot and Willis [13], Brown and Korringa [14], Rice and Cleary [15], Zimmerman [16], Berryman [17], Detournay and Cheng [18], Vardoulakis and Sulem [19], Coussy [20]. This framework is recalled here in a comprehensive manner in order to clarify the mathematical and physical significance of the different poromechanical parameters and effective stress coefficients which are evaluated in our experimental program.

## 2.1 Poroelastic formulation

A fluid saturated porous material can be seen as a mixture of two phases: a solid phase and a fluid phase. The solid phase may be itself made up of several constituents. The porosity $\phi$ is defined as the ratio of the volume of the porous space $V_\phi$ to the total volume $V$ in the actual (deformed) state.

$$\phi = \frac{V_\phi}{V} \tag{1}$$

For a saturated material the volume of the pore space is equal to the volume of the fluid phase. We consider a saturated sample under an isotropic state of stress $\sigma$ (positive in compression). We choose two independent variables for characterizing the volumetric behaviour of a porous material: the pore pressure $p_f$ and the differential pressure $\sigma_d$ which is equivalent to Terzaghi effective stress.

$$\sigma_d = \sigma - p_f \tag{2}$$

The expression of the variations of the total volume $V$ and of the pore volume $V_\phi$ introduces four parameters presented below [14]:

$$\frac{dV}{V} = \frac{1}{V}\left(\frac{\partial V}{\partial \sigma_d}\right)_{p_f} d\sigma_d + \frac{1}{V}\left(\frac{\partial V}{\partial p_f}\right)_{\sigma_d} dp_f$$

$$\frac{dV_\phi}{V_\phi} = \frac{1}{V_\phi}\left(\frac{\partial V_\phi}{\partial \sigma_d}\right)_{p_f} d\sigma_d + \frac{1}{V_\phi}\left(\frac{\partial V_\phi}{\partial p_f}\right)_{\sigma_d} dp_f \tag{3}$$

$$\frac{1}{K_d} = -\frac{1}{V}\left(\frac{\partial V}{\partial \sigma_d}\right)_{p_f} \quad , \quad \frac{1}{K_p} = -\frac{1}{V_\phi}\left(\frac{\partial V_\phi}{\partial \sigma_d}\right)_{p_f} \tag{4}$$

$$\frac{1}{K_s} = -\frac{1}{V}\left(\frac{\partial V}{\partial p_f}\right)_{\sigma_d} \quad , \quad \frac{1}{K_\phi} = -\frac{1}{V_\phi}\left(\frac{\partial V_\phi}{\partial p_f}\right)_{\sigma_d} \tag{5}$$

Equation (4) corresponds to a drained isotropic compression test in which the pore pressure is controlled to remain constant in the sample. The variations of the total volume of the sample $V$ and of the volume of the pore space $V_\phi$ with respect to the applied confining pressure give the drained bulk modulus $K_d$ and the modulus $K_p$. Equation (5) corresponds to the so-called unjacketed compression test, in which equal increments of confining pressure and pore pressure are simultaneously applied to the sample, as if the sample was submerged, without a jacket, into a fluid under the pressure $p_f$. The differential pressure $\sigma_d$ in this condition remains constant. The variation of the volume of the sample with respect to the applied pressure gives the unjacketed modulus $K_s$. The variation of the pore volume of the sample in this test, evaluated from the quantity of fluid exchanged between the sample and the pore pressure generator could in principle give the modulus $K_\phi$. However experimental evaluation of this parameter is very difficult as the volume of the exchanged fluid has to be corrected for the effect of fluid compressibility, and also for the effect of the deformations of the pore pressure generator and drainage system in order to access to the variation of the pore volume of the sample. In the case of a porous material which is homogeneous and isotropic at the micro-scale, the sample would deform in an unjacketed test as if all the pores were filled with the solid component. The skeleton and the solid component experience a uniform volumetric strain with no change of the porosity [18]. For such a material $K_s = K_\phi = K_m$, where $K_m$ is the bulk modulus of the single solid constituent of the porous material. In the case of a porous material which is composed of two or more solids and therefore is heterogeneous at the micro-scale, the unjacketed modulus $K_s$ is some weighted average of the bulk moduli of solid constituents [17]. What this average should be is generally unknown, however, in Ghabezloo and Sulem [21] the unjacketed modulus of Rothbach sandstone was evaluated using Hill's [22] average formula and was in good accordance with the experimentally evaluated modulus. The modulus $K_\phi$ for such a material has a complicated dependence on the material properties. Generally it is not bounded by the elastic moduli of the solid components and can even have a negative sign if the bulk moduli of the individual solid components are greatly different one from another [23][24].

Using Betti's reciprocal theorem one obtains the following relation between the elastic moduli [14][25]:

$$\frac{1}{K_p} = \frac{1}{\phi}\left(\frac{1}{K_d} - \frac{1}{K_s}\right) \tag{6}$$

Using equation (6) the number of required moduli to characterize the volumetric behaviour of a porous material is reduced to three. Using the four moduli defined in equations (4) and (5), the variations of the total volume $V$ and of the pore volume $V_\phi$ (equation (3)) can be rewritten as follows:

$$\frac{dV}{V} = -\frac{d\sigma_d}{K_d} - \frac{dp_f}{K_s}$$
$$\frac{dV_\phi}{V_\phi} = -\frac{d\sigma_d}{K_p} - \frac{dp_f}{K_\phi} \tag{7}$$

The incremental volumetric strain $d\varepsilon = -dV/V$ is thus expressed as:

$$d\varepsilon = \frac{d\sigma_d}{K_d} + \frac{dp_f}{K_s} \tag{8}$$

Using the definition of the porosity presented in equation (1), the following equation is obtained for the variation of the porosity:

$$\frac{d\phi}{\phi} = \frac{dV_\phi}{V_\phi} - \frac{dV}{V} \tag{9}$$

Replacing equation (7) and then equation (6) in equation (9), the expression of the variation of porosity is found:

$$\frac{d\phi}{\phi} = -\frac{1}{\phi}\left(\frac{1-\phi}{K_d} - \frac{1}{K_s}\right)d\sigma_d + \left(\frac{1}{K_s} - \frac{1}{K_\phi}\right)dp_f \tag{10}$$

The undrained condition is defined as a condition in which the mass of the fluid phase is constant ($dm_f = 0$). Under this condition we choose two different independent variables: The total stress $\sigma$ and the fluid mass $m_f$. The measured quantities are the total volume $V$ and the pore pressure $p_f$. Writing the expression of the variation of these quantities, we can define two new parameters to describe the response of the porous material in undrained condition:

$$\frac{1}{K_u} = -\frac{1}{V}\left(\frac{\partial V}{\partial \sigma}\right)_{m_f} \quad , \quad B = \left(\frac{\partial p_f}{\partial \sigma}\right)_{m_f} \tag{11}$$

The parameter $K_u$ is the undrained bulk modulus and $B$ is the so-called Skempton coefficient [26]. As the fluid mass in the undrained condition is constant, the variation of the volume of the fluid is given by the fluid compression modulus $K_f$ and the variation of the pore pressure:

$$\frac{dV_\phi}{V_\phi} = -\frac{dp_f}{K_f} \tag{12}$$

Replacing equation (12) in equation (7) and using equations (2) and (6) the expression of the Skempton coefficient $B$ (equation (11)) is obtained:

$$B = \frac{(1/K_d - 1/K_s)}{(1/K_d - 1/K_s) + \phi(1/K_f - 1/K_\phi)} \tag{13}$$

The variation of the total volume in undrained condition is given by the undrained bulk modulus $K_u$ as presented in equation (11). Replacing $d\varepsilon = -dV/V = d\sigma/K_u$, $d\sigma_d = d\sigma - dp_f$ and $dp_f = Bd\sigma$ in equation (8), the following expression is found for the undrained bulk modulus $K_u$:

$$K_u = \frac{K_d}{1 - B(1 - K_d/K_s)} \tag{14}$$

Biot modulus $M$ can be expressed in terms of the elastic moduli defined in equations (4) and (5) [18]:

$$\frac{1}{M} = \frac{K_d}{K_s}\left(\frac{1}{K_d} - \frac{1}{K_s}\right) + \phi\left(\frac{1}{K_f} - \frac{1}{K_\phi}\right) \tag{15}$$

As mentioned earlier, the modulus $K_\phi$ is very difficult to measure experimentally. On the other hand, the experimental evaluation of the poroelastic parameters $K_d$, $K_s$, $K_u$ and $B$ is more common, so that using these moduli that can be measured independently and equations

(13) and (14) one can find four different expressions for indirect evaluation of the parameter $K_\phi$.

$$\frac{1}{K_\phi} = \begin{cases} \dfrac{1}{K_f} - \dfrac{(1/K_d - 1/K_s)(1/K_u - 1/K_s)}{\phi(1/K_d - 1/K_u)} & \text{(a)} \\ \dfrac{1}{K_f} - \dfrac{(1-B)(1/K_d - 1/K_s)}{\phi B} & \text{(b)} \\ \dfrac{1}{K_f} - \dfrac{1/K_u - 1/K_s}{\phi B} & \text{(c)} \\ \dfrac{1}{K_f} - \dfrac{(1-B)(1/K_d - 1/K_u)}{\phi B^2} & \text{(d)} \end{cases} \quad (16)$$

## 2.2 Effective stress concept

The concept of effective stress was first introduced by Terzaghi [27] who defined it as the difference between the total stress and the pore pressure, and attributed all measurable effects of a change in stress exclusively to changes in the effective stress.

More generally, the effective stress $\sigma' = \sigma'(\sigma, p_f)$ can be defined as a stress quantity which can be used as a single variable to express the stress dependency of a property $Q$ of a porous material. Reducing the number of independent variables from two to one using the concept of effective stress greatly simplifies the analysis of total stress and pore pressure dependency of porous material properties.

$$Q = Q(\sigma, p_f) = Q(\sigma') \quad (17)$$

If the total stress and the pore pressure vary in such a way that the effective stress remains constant, then no variation in the corresponding property $Q$ is expected. Thus the expression of the effective stress can be obtained from the evaluation of the isolines of $Q(\sigma, p_f)$.

Since different material properties may depend on total stress and pore pressure in different ways, there is not a unique effective stress which would be appropriate for all properties of the material, and consequently different effective stress expressions should be defined for the different properties.

Let us now consider $Q$ as a physical property of the porous material which is a function of the differential pressure and of the pore pressure $Q = Q(\sigma_d, p_f)$. Assuming that this function is smooth enough so that its derivatives can be defined, the incremental variation of $Q$ can be written in the following form:

$$dQ = \frac{\partial Q}{\partial \sigma_d} d\sigma_d + \frac{\partial Q}{\partial p_f} dp_f \tag{18}$$

Introducing the definition of $\sigma_d$ (equation (2)) in equation (18) we obtain:

$$dQ = \frac{\partial Q}{\partial \sigma_d}\left[d\sigma - \left(1 - \frac{\partial Q/\partial p_f}{\partial Q/\partial \sigma_d}\right)dp_f\right] \tag{19}$$

The above expression shows that the variation of the property $Q$ can be written as a function of a single incremental quantity $d\sigma'$:

$$dQ = \frac{\partial Q}{\partial \sigma_d} d\sigma' \tag{20}$$

where $d\sigma'$ is defined by

$$d\sigma' = d\sigma - n_Q dp_f \tag{21}$$

with

$$n_Q(\sigma_d, p_f) = 1 - \frac{\partial Q/\partial p_f}{\partial Q/\partial \sigma_d} \tag{22}$$

The isolines of $Q$ are obtained by the integration of the following differential equation:

$$d\sigma' = 0 \tag{23}$$

In the close vicinity of a given state of stress $(\sigma_d, p_f)$, the isolines are generally approximated with parallel straight lines (e.g. Bernabé [28]) which is equivalent to the assumption that $n_Q$ is a constant. Under this assumption, equation (21) can be easily integrated and a linear expression for the effective stress is obtained:

$$\sigma' = \sigma - n_Q p_f \tag{24}$$

The above expression is the most common form of the effective stress as used in the mechanics of porous media. The effective stress coefficient $n_Q$ is equal to one in Terzaghi's definition which means that the total stress and the pore pressure have similar, but inverse, effects on the variation of the property $Q$. While a value of the coefficient $n_Q$ smaller (respectively greater) than unity means that the effect of pore pressure change on the variation of the property $Q$ is less (respectively more) important than the effect of a change in total stress. The expression of the effective stress coefficient presented in equation (22) was first presented by Todd and Simmons [29].

Zimmerman [16] and Berryman [17] have derived general effective stress rules for various physical properties of rocks and presented the expressions of the effective stress coefficients $n_Q$ corresponding to different physical properties together with some bounds and general relations among these coefficients.

If we replace the quantity $Q$ in equation (22) by the incremental volumetric strain $d\varepsilon = -dV/V$ and using the definitions of bulk moduli presented in equations (4) and (5), the expression of the Biot [13] effective stress coefficient $\alpha$ for the total volume change of a porous material is retrieved:

$$\alpha = 1 - \frac{K_d}{K_s} \tag{25}$$

Similarly, replacing $Q$ by $-dV_\phi/V_\phi$ and using equations (4) and (5), the effective stress coefficient $\beta$ corresponding to the variation of pore volume $V_\phi$ is found:

$$\beta = 1 - \frac{K_p}{K_\phi} = 1 - \frac{\phi}{K_\phi(1/K_d - 1/K_s)} \tag{26}$$

The effective stress coefficient $\chi$ for the variation of porosity $\phi$ can be obtained by re-writing equation (10) using equation (2):

$$d\phi = -\left(\frac{1-\phi}{K_d} - \frac{1}{K_s}\right)(d\sigma - \chi dp_f) \tag{27}$$

with

$$\chi = 1 - \frac{\phi(1/K_\phi - 1/K_s)}{(1-\phi)/K_d - 1/K_s} \tag{28}$$

Using equation (1) the following expression can be obtained for the variations of the volume of the solid phase $V_s = V - V_\phi$:

$$-\frac{dV_s}{V_s} = \frac{dV}{V} - \frac{d\phi}{1-\phi} \tag{29}$$

Introducing equations (8) and (10) in equation (29) and re-writing the resulted equation, the following expression is obtained for the effective stress coefficient $\kappa$ for the variations of the volume of solid phase:

$$\kappa = \frac{\phi K_s}{K_\phi} \tag{30}$$

Berryman [17] has shown that the above set of effective stress coefficients for the variations of different physical properties of porous materials verifies the following inequality:

$$\kappa \leq \alpha \leq \beta \leq \chi \tag{31}$$

Berryman underlines that the inequality $\alpha \leq \chi$ is obtained by considering the empirical inequality $\phi \leq \alpha$.

### 2.2.1 Effective stress for the drained bulk modulus

Using equation (2), equation (8) can be re-written in the following form:

$$d\varepsilon = \frac{d\sigma}{K_d} + \left(\frac{1}{K_s} - \frac{1}{K_d}\right)dp_f \tag{32}$$

For hyperelastic materials, the existence of an elastic potential guaranties that the volumetric strain $\varepsilon$ is independent of the loading path. This condition is fulfilled when the Euler condition of integrability in equation (32) is verified:

$$\frac{\partial (1/K_d)}{\partial p_f} = \frac{\partial (1/K_s - 1/K_d)}{\partial \sigma} \tag{33}$$

Considering now the drained compressibility $c_d = 1/K_d$ as a function of $\sigma$ and $p_f$, the variation of $c_d$ can be written in the following form:

$$dc_d = \frac{\partial c_d}{\partial \sigma} d\sigma + \frac{\partial c_d}{\partial p_f} dp_f \tag{34}$$

This equation can be re-written in the following form:

$$dc_d = \frac{\partial c_d}{\partial \sigma} \left( d\sigma + \frac{\partial c_d / \partial p_f}{\partial c_d / \partial \sigma} dp_f \right) \tag{35}$$

The effective stress coefficient $\theta$ corresponding to the variations of the drained bulk modulus can be defined as:

$$\theta = -\frac{\partial c_d / \partial p_f}{\partial c_d / \partial \sigma} \tag{36}$$

Inserting equation (33) in equation (36) and replacing $c_d$ with $1/K_d$, the following expression is found:

$$\theta = 1 - \frac{\partial (1/K_s)/\partial \sigma}{\partial (1/K_d)/\partial \sigma} \tag{37}$$

We can see that for a material with a constant unjacketed modulus $K_s$, the effective stress coefficient $\theta$ is equal to one and the drained bulk modulus is a function of Terzaghi effective stress $K_d = f(\sigma_d)$. Various theoretical demonstrations of this statement can be found in Zimmerman [16], Berryman [17], Coussy [20] and Gurevich [30] and from micromechanical considerations in Dormieux et al. [31]. Experimental verification is presented by Boutéca et al. [32] for two sandstones.

Coyner [33] has performed experimental measurements of the variations of the unjacketed and of the drained bulk moduli of different rocks (sandstones, limestones and granites), with a range of pressures up to 100MPa. The results showed that the maximum variation of the unjacketed modulus for the range of applied pressures is about 10%. We can evaluate the effective stress coefficients $\theta$ using the experimental data of Coyner [33] and we find numerical values very close to one, between 0.95 and 1.0. These experimental results confirm that, even for a porous material which is heterogeneous at the micro-scale and for which a small variation of the unjacketed compression modulus with applied stress is expected, this variation is insignificant regarding the effective stress coefficients $\theta$. Thus, for most practical applications, the dependence of the bulk modulus with *Terzaghi* effective stress is an acceptable assumption.

## 2.3 Tangent and secant moduli

The variations of the volume of a porous material can be expressed in terms of either tangent or secant moduli. The definitions are presented in Figure (1) and equations (4) and (5). From Figure (1) we can write the following relationships between the secant and tangent moduli:

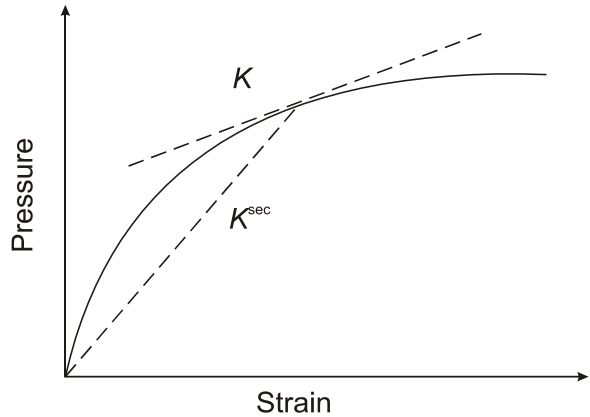

**Figure 1-Definitions of the tangent and secant moduli**

$$\frac{1}{K_d^{\text{sec}}} = \frac{1}{\sigma_d}\int \frac{d\sigma_d}{K_d} \tag{38}$$

$$\frac{1}{K_s^{\text{sec}}} = \frac{1}{p_f}\int \frac{dp_f}{K_s} \tag{39}$$

The secant modulus can be interpreted as an average value of the tangent modulus over a range of pressure. If the drained bulk modulus is a function of the Terzaghi effective stress and the unjacketed modulus is constant, which is the case for most geomaterials as mentioned in section 2.2.1, then equation (8) can be integrated and the volumetric strain can be written using the secant properties. This assumption is similar to the one used by Zimmerman *et al.* [25] and will be justified in section 3 for the studied material.

$$\varepsilon = \frac{\sigma_d}{K_d^{\text{sec}}} + \frac{p_f}{K_s^{\text{sec}}} \tag{40}$$

We can re-write this equation, using equation (2), to find the expression of the secant Biot effective stress coefficient:

$$\varepsilon = \frac{1}{K_d^{\text{sec}}}\left(\sigma - \alpha^{\text{sec}} p_f\right) \tag{41}$$

with

$$\alpha^{\text{sec}} = 1 - \frac{K_d^{\text{sec}}}{K_s^{\text{sec}}} \tag{42}$$

The expression of Biot effective stress $\sigma'_b$ is obtained from equation (41):

$$\sigma'_b = \sigma - \alpha^{\text{sec}} p_f \tag{43}$$

## 2.4 Cement paste porosity for poromechanical formulations

The size of pores in the microstructure of hardened cement paste covers an impressive range, from nanometre-sized gel pores, to micro-metre sized capillary pores and millimetre-sized air voids [34]. From the different microstructural models of C-S-H, it can be seen that a part of the water in the pore structure of cement paste is interlayer structural water. Feldman and Sereda [35] propose a model for multilayer structure of C-S-H that postulates the existence of interlayer spaces containing strongly adsorbed water. According to Feldman [36], the interlayer water behaves as a solid bridge between the layers and consequently the interlayer space can not be included in the porosity and the interlayer water must be regarded as a part of the solid structure of hydrated cement paste. Feldman's experiments show that the interlayer water evaporates at very low relative humidity, below 11%. This can be seen also in Jennings' [37][38] microstructural model of C-S-H, in which the amorphous colloidal structure of the C-S-H is organized in elements, called 'globules'. In Jennings' [38] model, the globule is composed of solid C-S-H sheets, intra-globule porosity and a monolayer of water on the surface. For relative humidities below 11%, a part of the water filling the intra-globule porosity is evaporated. In a porosity measurement procedure in which the sample is dried at 105°C until a constant mass is achieved, a part of the interlayer water is evaporated and is thus included in the measured porosity [34][39]. Consequently, the porosity measured by oven-drying at 105°C should not be used in poromechanical formulation, as it includes a part of the interlayer porosity. The free-water porosity is defined and measured by equilibrating the cement paste at 11% relative humidity [34][39]. Some measurements of the free-water porosity, which is obviously smaller that the total porosity measured by drying at 105°C, from Feldman [40] are presented by Jennings *et al*. [34]. From the above discussion it appears that the free-water porosity is the cement paste porosity that should be used in the poromechanical formulations. This is in accordance with the assumption made by Ulm *et al*. [1] in their multi-scale microstructural model for the evaluation of the poromechanical properties of hardened cement paste. These authors use the Jennings' [37] model and exclude the intra-globule porosity from the cement paste porosity used in the calculations. According to Ulm *et al*. [1], considering the characteristic size of the interlayer space, which is less than ten water molecules in size, it is recognized that the water in this space can not be regarded as a bulk water phase.

Taylor [39] (in his Fig. 8.5) presents the values of total and free-water porosity derived from calculated phase compositions of mature ordinary cement paste with varying w/c ratio. The comparison of the calculated free-water porosity with the porosity measured by mercury intrusion shows that for w/c ratios smaller than 0.5, the free-water porosity can be approximated by the mercury porosity. For higher w/c ratios, the free-water porosity is somewhat higher than the mercury porosity.

## 3 Experimental program

In order to evaluate the parameters defined in the above section, drained and undrained isotropic compression tests as well as unjacketed tests have been performed on a hardened cement paste. The effective stress law corresponding to the variations of the drained bulk modulus is studied by performing the drained compression tests with different constant pore pressures.

### 3.1 Sample preparation

A class G oil well cement was used to prepare the cement paste with a water to cement ratio $w/c = 0.44$. Two additives, a dispersant and an anti-foaming agent were used in the paste.

The fresh paste was conserved in 14cm cubic moulds during four days in lime saturated water at 90°C temperature. After this period, the temperature was reduced gradually to prevent the cracking of the blocs due to a sudden temperature change. Then, the blocs were cored and cut to obtain cylindrical samples with 38mm diameter and 76mm length. The two ends of the cylindrical samples were rectified to obtain horizontal surfaces. After the preparation of the samples, the geometry and the weight of the samples were measured. To insure the homogeneity and the integrity of the samples, measurements of wave velocity and dynamic elastic modulus were performed on all of them. These measurements were performed at ambient temperature.

After the sample preparation phase, the samples have been submerged in a fluid which is neutral regarding to the pore fluid of the cement paste to prevent chemical reactions during the period of curing. The samples have been cured during at least three months at 90°C in the neutral fluid for which pH=13 before performing the mechanical loading tests. Before testing, the temperature of each sample was reduced slowly to prevent any thermal cracking.

The porosity of the samples is studied by two methods: oven drying and mercury intrusion porosimetry. The total porosity is measured by drying the samples at 105°C until a constant mass is achieved, and an average value equal to $\phi = 0.35$ is obtained. As mentioned in section 2.4, this porosity includes a part of the interlayer water of the cement paste. The mercury intrusion porosimetry is performed on the samples which are dried before the tests with the freeze-drying technique using liquid nitrogen which is, according to Gallé [41], the most suitable drying procedure to investigate the pore structure of cement-based materials. With a maximum intruding pressure of 200MPa the average mercury porosity of the samples is obtained equal to $\phi = 0.26$ (Figure 2). Using the Washburn-Laplace equation ($P = -4\gamma\cos\theta/d$) and assuming a contact angle $\theta$ of 130° and a surface tension of mercury $\gamma$ of 0.483N/m from Ref. [39], the maximum intruding pressure $P$ of 200MPa corresponds to a minimum pore diameter of about 6nm. Based on the discussions of section 2.4, this value of $\phi = 0.26$ will be used as an approximation of the free-water porosity of the studied cement paste in the following poromechanical formulations.

## 3.2 *Experimental setting*

The triaxial cell used in this study can sustain a confining pressure up to 60MPa. It contains a system of hydraulic self-compensated piston. The loading piston is then equilibrated during the confining pressure build up and directly applies the deviatoric stress. The axial and radial strains are measured directly on the sample inside the cell with two axial transducers and four radial ones of LVDT type. The confining pressure is applied by a servo controlled high pressure generator. Hydraulic oil is used as confining fluid. The pore pressure is applied by another servo-controlled pressure generator with possible control of fluid volume or pore pressure. Further details on the testing system can be found in Sulem and Ouffroukh [42] and Ghabezloo and Sulem [21].

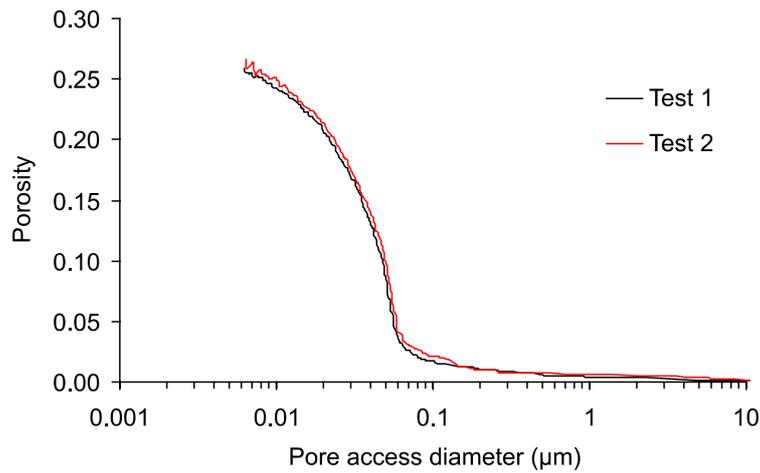

**Figure 2-Results of porosity measurements by mercury intrusion**

## 3.3 Sample saturation and loading rate

During the period of curing, the samples are stored inside the saturating fluid at 90°C for at least three months. After this period they can be considered as completely saturated. The process of installing the sample inside triaxial cell may cause a partial de-saturation of the sample. For this reason a seven days re-saturation phase is performed inside the triaxial cell. During this period, the sample is maintained under a confining pressure equal to 1.2MPa and a back fluid pressure equal to 1.0MPa is applied to the sample while the volume of the fluid injected in the sample is monitored. To prevent the chemical reactions during the test, the same neutral fluid which is used during the period of curing is used in the pore pressure generator as the saturating fluid.

For a drained test, the loading rate should be slow enough to ensure the drainage of the generated excess pore pressure in the sample. This loading rate is a function of the permeability of the tested material and of the drainage length of the sample. In order to determine the appropriate loading rate, a preliminary study was done in which the sample was in simple drainage condition. The pore pressure is maintained constant in one side of the sample. The valve situated at the other side was closed and the pore pressure was monitored using a pore pressure transducer. The confining pressure with different loading rates (0.1MPa/min, 0.075MPa/min, 0.05MPa/min, 0.02MPa/min) was applied and we observed that by using a loading rate equal to 0.02MPa/min no excess pore pressure was measured in the sample. Consequently, in order to ensure a good drainage of the sample during the tests, a loading rate equal to 0.025MPa/min is used with a double drainage system, in which the pore pressure is controlled at both sides of the sample. For undrained compression tests, a loading rate equal to 0.10MPa/min is used.

## 3.4 Unjacketed test

The unjacketed test is performed by applying simultaneous and equal variations of confining pressure and pore pressure. A small difference between the confining pressure and pore pressure equal to 0.5MPa is maintained inside the sample in order to avoid any leakage of the pore fluid between the sample and the membrane. A loading-unloading cycle is applied and the unjacketed modulus is calculated for the unloading part of the curve. The result is presented in the Figure (3) and shows that the unjacketed modulus does not vary with the pressure up to 50MPa and is equal to 21GPa.

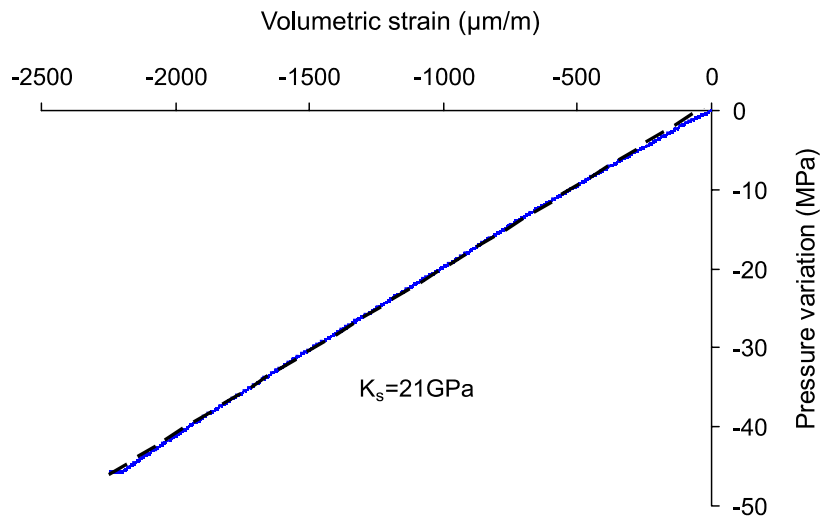

**Figure 3- Unjacketed test, pressure-volumetric strain response in unloading**

Considering that the unjacketed modulus does not vary with the applied pressure, using equation (37) we can conclude that the effective stress coefficient corresponding to the drained bulk modulus of the hardened cement paste is equal to one and that the bulk modulus varies with Terzaghi effective stress. This conclusion can be verified using the results of the drained compression tests and the variations of the measured modulus with confining pressure and pore pressure.

## 3.5 Drained compression tests

Five drained isotropic compression tests have been performed. In each test, some unloading-reloading cycles have been performed in order to evaluate the secant bulk modulus of the hardened cement paste. Three tests are performed with a constant pore pressure equal to 1.0MPa. Two other tests are performed using pore pressures equal respectively to 23.5MPa and 47.0MPa. The loading paths are presented in the Figure (4). Table (1) presents the details of the tests and of the loading cycles.

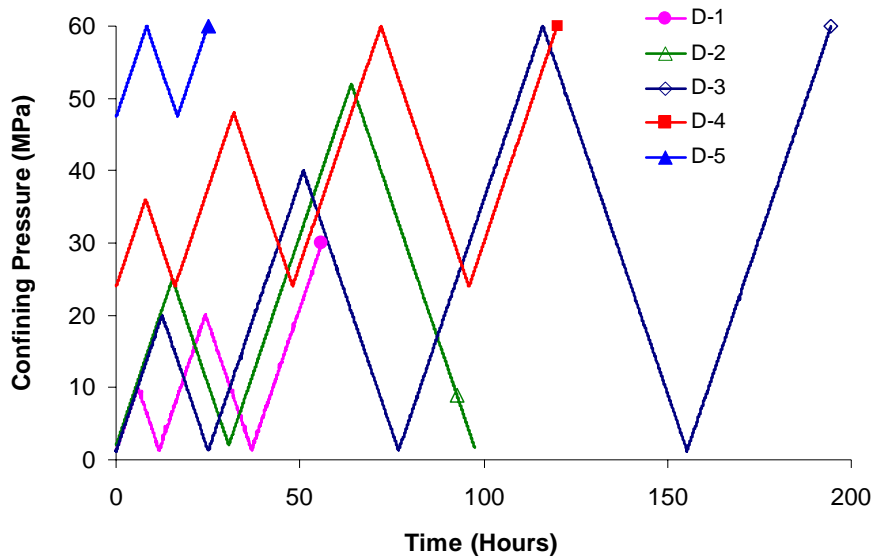

**Figure 4- Loading paths of drained isotropic compression tests (D-1, D-2, D-3 with constant applied pore pressure of 1.0MPa, D-4 with constant applied pore pressure of 23.5MPa, D-5 with constant applied pore pressure of 47.0MPa)**

**Table 1- Details of the performed drained isotropic compression tests**

| Test | Pore pressure (MPa) | Cycle n°1 (MPa) | Cycle n°2 (MPa) | Cycle n°3 (MPa) |
|---|---|---|---|---|
| D-1 | 1.0 | 10 | 20 | - |
| D-2 | 1.0 | 25 | 52 | - |
| D-3 | 1.0 | 20 | 40 | 60 |
| D-4 | 23.5 | 36 | 48 | 60 |
| D-5 | 47.0 | 60 | - | - |

The stress-strain curves obtained from the isotropic drained compression tests are presented in Figure (5). We can see the strongly nonlinear response of the hardened cement paste under isotropic compression and the existence of the permanent strains. Due to the slow rate of the loading, a part of the measured strains, especially at high stresses, may be due to creep. This point can easily be observed for example in the last unloading parts of D-2 and D-3 tests, where despite of the diminution of the confining pressure, the volumetric strain continues to increase. These effects are ignored in this paper and will be discussed in a future publication. The drained bulk modulus is calculated for each unloading-reloading cycle. The method of evaluation of the modulus is presented in the Figure (6). A linear fitting of the whole unloading-reloading curve is performed and the slope of the obtained line is considered as an average secant modulus corresponding to the stress level at the beginning of the unloading phase.

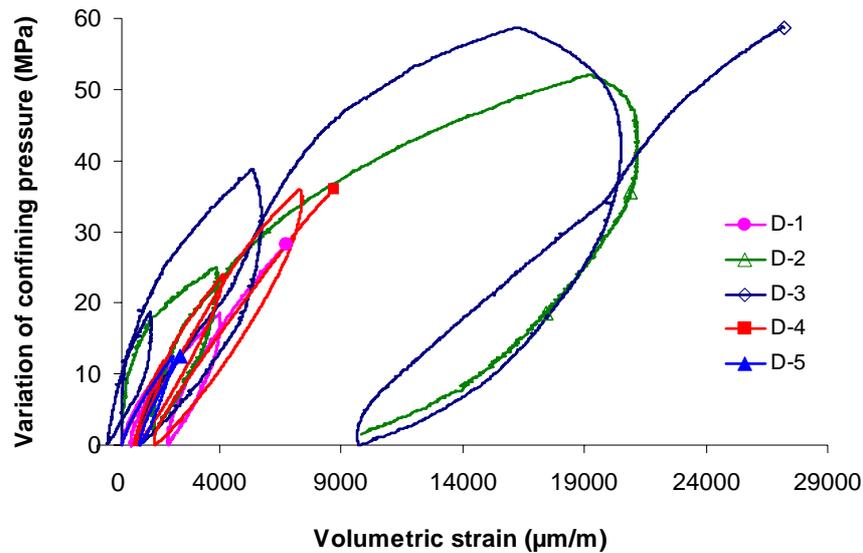

**Figure 5- Drained isotropic compression tests, Pressure-volumetric strain curves**

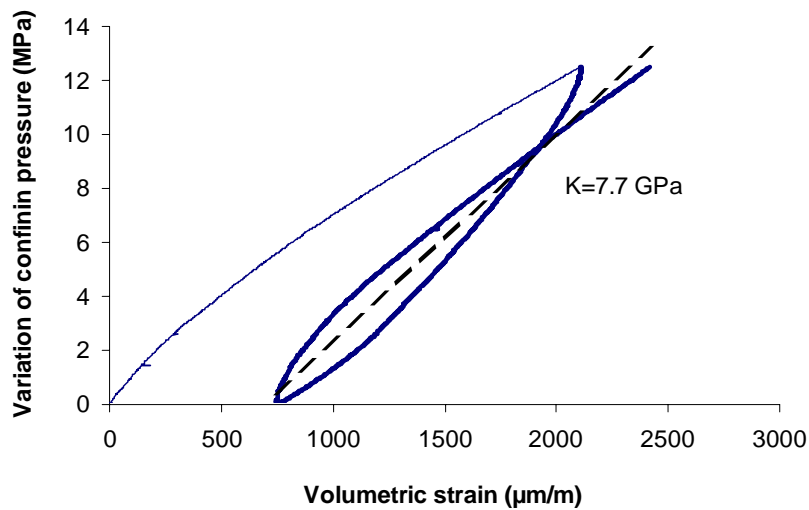

**Figure 6- Evaluation of the secant bulk modulus in an unloading-reloading cycle**

Figure (7) shows the values of measured bulk modulus for different confining pressures and different pore pressures. Looking at the points corresponding to the tests performed with the pore pressure equal to 1.0MPa and 23.5MPa we observe a reduction of the bulk modulus with the confining pressure. This behaviour and the phenomenon of degradation of the hardened cement paste and reduction of the bulk modulus under isotropic loading differs from the one commonly observed in granular porous rocks. For these rocks the bulk modulus generally increases under isotropic stress due to the closure of the pre-existing microcracks and compaction of the rock matrix. This mechanism of degradation of the bulk modulus and the highly non-linear behaviour of hardened cement paste under isotropic loading will be studied further in this paper by comparison of microphotographs of the samples before and after the compression tests.

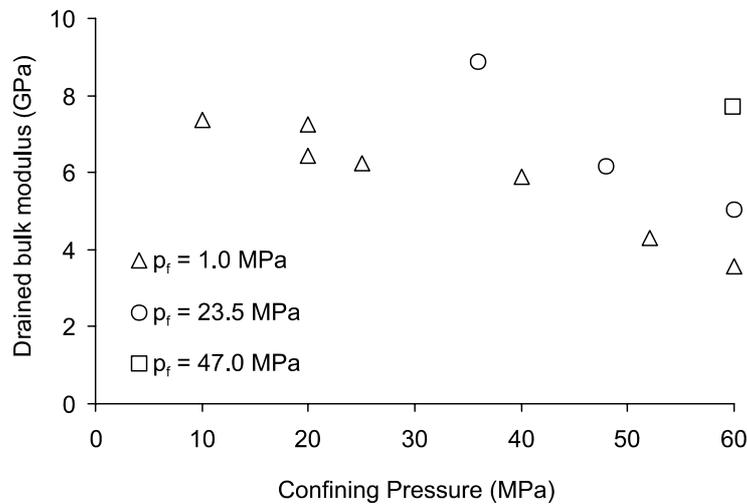

**Figure 7- Variation of the drained bulk modulus with the confining pressure at various imposed pore pressures**

It can be seen in Figure (7) that at a given confining pressure, the bulk modulus increases with increase of the pore pressure. This graph shows the simultaneous effects of the confining pressure and of the pore pressure on the degradation of the hardened cement paste and on the variation of the bulk modulus, and highlights the necessity of finding a single stress variable (effective stress) to describe this phenomenon.

As mentioned above, as the unjacketed modulus was found to be constant, the drained bulk modulus varies with Terzaghi effective stress. This statement can be experimentally verified here by using the results of drained compression tests. Figure (8) shows the evaluated drained bulk moduli as a function of Terzaghi effective stress calculated for each data point using the corresponding values of confining pressure and pore pressure. We can see the good compatibility of the drained bulk moduli measured with different pore pressures when presented as a function of Terzaghi effective stress. The experimental results thus confirm the above conclusion and the following expression is obtained for the drained bulk modulus of hardened cement paste as a function of Terzaghi effective stress:

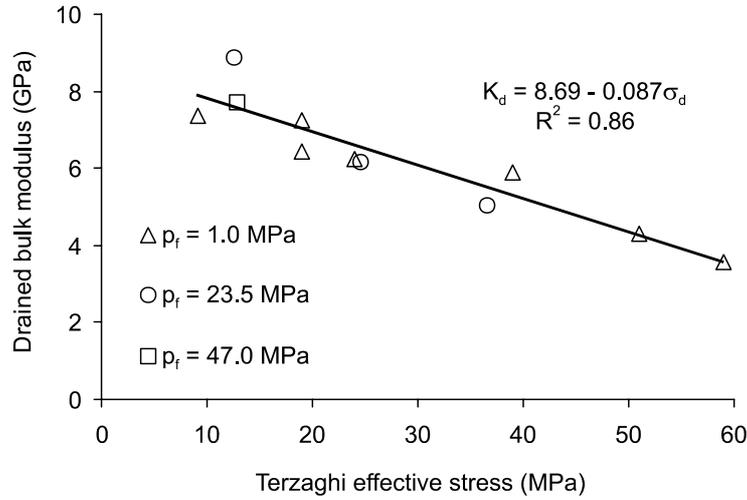

**Figure 8- Variation of the drained bulk modulus with Terzaghi effective stress**

$$K_d^{\text{sec}} = 8.69 - 0.087\sigma_d \quad (K_d^{\text{sec}} \text{ in GPa and } \sigma_d \text{ in MPa}) \tag{44}$$

Equation (44) shows that the bulk modulus at zero effective stress is about 8.7GPa. On the other hand, the bulk modulus can be estimated from a simple relationship used by Heukamp *et al.* [11] from [43] between $K_s$, $K_d$ and $\phi$:

$$K_d = K_s (1-\phi)^3 \tag{45}$$

Replacing $K_s = 21\text{GPa}$ and $\phi = 0.26$ in equation (45), the drained bulk modulus is found equal to 8.5GPa which has a very good compatibility with the measured values.

A similar expression in the form $E = E_0 (1-\phi)^3$ was used by Helmuth and Turk [44] to approximate the variations of the Young modulus of the hardened cement paste with the porosity. The parameter $E_0$ is the value of Young modulus at zero porosity limit and was found equal to 29GPa. The simple cubic law for the relation between the porosity and the bulk modulus (equation (45)) can be obtained from structural mechanics approaches at the micro-scale as derived for cellular solids [45]. This relation verifies the Hashin-Shtrikman [46] bounds and is very close to the modulus obtained from a self-consistent homogenisation scheme for porosities smaller that 0.45 (Figure (9)).

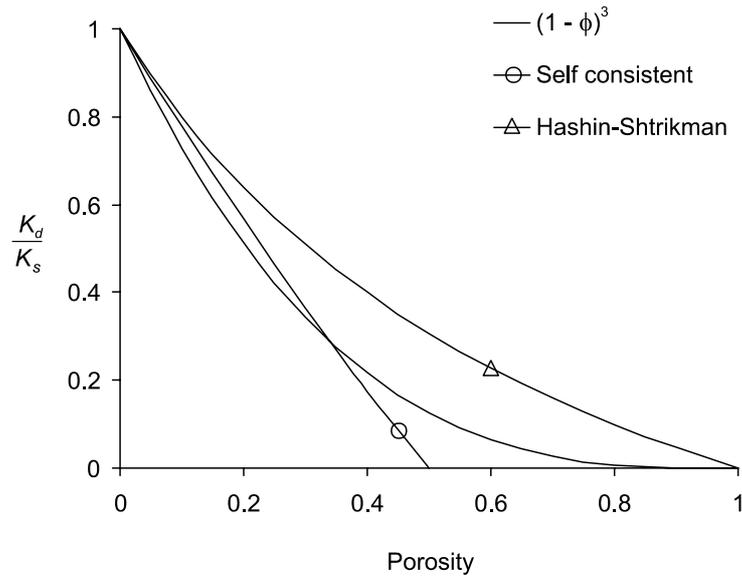

**Figure 9- Comparison of equation (45) with Hashin-Shtrikman bound and self consistent homogenisation scheme**

The expression of the drained bulk modulus presented in equation (44) and the value of the unjacketed modulus of the hardened cement paste (21GPa), enable us to determine the expression of the Biot effective stress coefficient presented in equation (42).

$$\alpha^{\text{sec}} = 0.586 + 0.004\sigma_d \qquad (\sigma_d \text{ in MPa}) \qquad (46)$$

Inserting equation (45) in equation (25) the following relation is found for the Biot effective stress coefficient:

$$\alpha = 1 - (1-\phi)^3 \qquad (47)$$

Replacing $\phi = 0.26$ in this relation, the Biot effective stress coefficient is found equal to 0.59 which is compatible with the result presented in equation (46).

## 3.6 Undrained compression tests

In order to determine the undrained poroelastic parameters of hardened cement paste, (undrained bulk modulus $K_u$ and Skempton coefficient $B$ defined in equation (11)), two undrained isotropic compression tests were performed with three unloading-reloading cycle in each test. Table (2) presents the details of the cycles and the initial conditions of the test. The loading paths are shown in Figure (10). The measured volumetric strains during the tests are presented in Figure (11). The undrained behaviour of hardened cement paste observed in these tests shows a lower non-linearity as compared to the results of drained compression tests presented in Figure (5). We can also observe that the permanent strains measured in the unloading parts are considerably less than the strains measured in the drained tests. The undrained bulk moduli are evaluated with the same method as presented above in Figure (6) for the drained tests. It is assumed here that the undrained parameters, $K_u$ and $B$, vary with Terzaghi effective stress. As it can be seen in equations (13) and (14), this assumption is verified in the case of an ideal porous material, $K_s = K_\phi = K_m$, if the solid and the fluid compression moduli, $K_m$ and $K_f$, are assumed constant. As shown by Zimmerman [16], for usual values of the moduli which appear in equations (13) and (14), the undrained parameters are almost completely insensitive to deviations of $K_\phi$ from $K_s$ and the results in the case of an inhomogeneous porous material differ only slightly from the one obtained for an ideal porous material. Moreover, this assumption is adopted and commonly used in different experimental studies [47][48][49][50]. The influence of this assumption on the analysis of the experimental results is studied later in this paper.

**Table 2- Details of the performed undrained isotropic compression tests**

| Test | Initial confining pressure (MPa) | Initial pore pressure (MPa) | Cycle n°1 (MPa) | Cycle n°2 (MPa) | Cycle n°3 (MPa) |
|---|---|---|---|---|---|
| ND-1 | 1.65 | 1.50 | 20 | 40 | 60 |
| ND-2 | 1.85 | 1.65 | 15 | 30 | 50 |

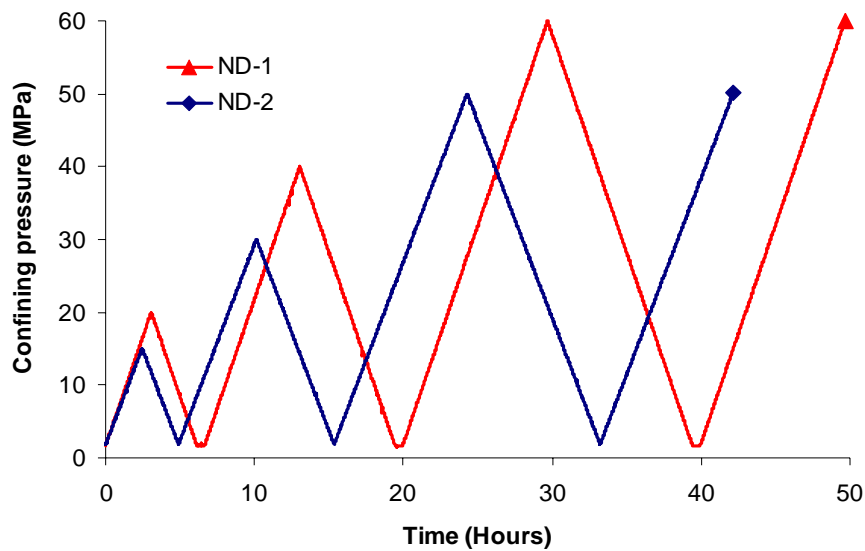

**Figure 10- Loading paths of the undrained isotropic compression tests**

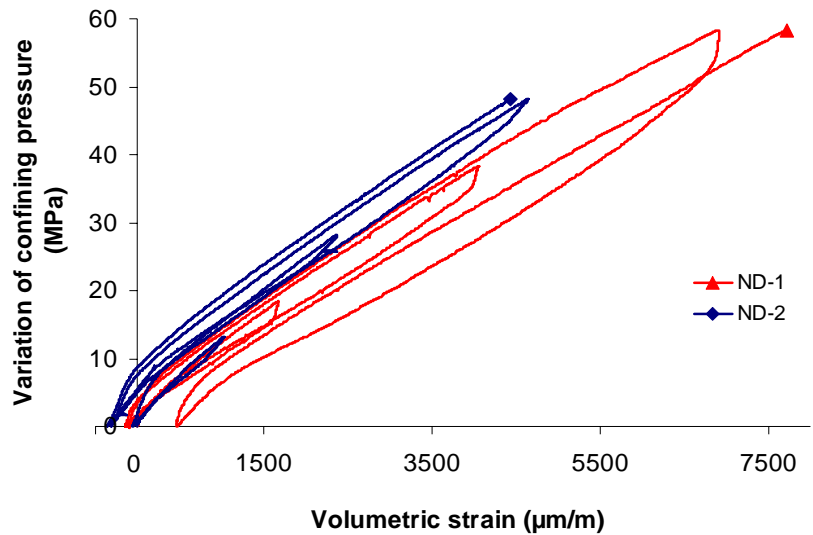

**Figure 11- Undrained isotropic compression tests, pressure-volumetric strain curves**

Figure (12) presents the undrained bulk moduli evaluated in unloading-reloading cycles. The corresponding Terzaghi effective stress is calculated for each point from the imposed confining pressure and the measured pore pressure. These data are fitted with a straight line:

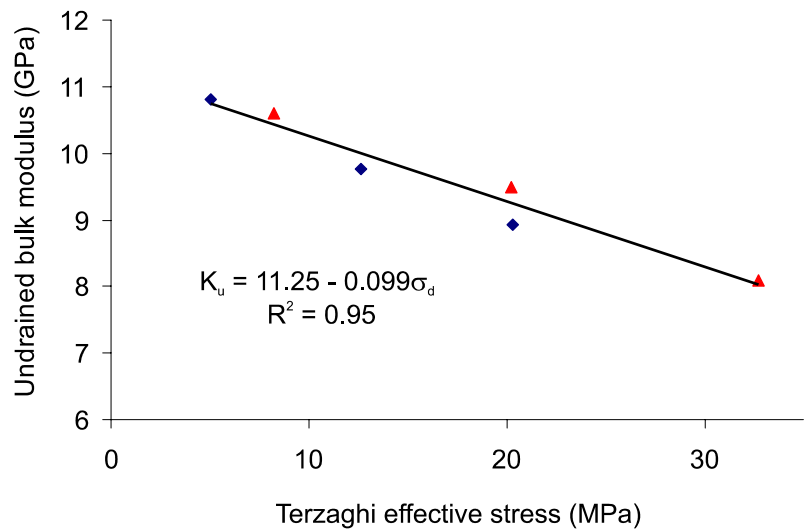

**Figure 12- Variation of the undrained bulk modulus with Terzaghi effective stress**

$$K_u^{sec} = 11.25 - 0.099\sigma_d \quad (K_u^{sec} \text{ in GPa and } \sigma_d \text{ in MPa}) \quad (48)$$

As expected, for an effective stress, the undrained bulk modulus calculated by equation (48) is greater than the drained bulk modulus calculated with equation (44). The mechanism of degradation of the undrained modulus is discussed further using the results of microscopic observations of the tested samples.

Figure (13) presents the variations of the pore pressure during the undrained loading with the confining pressure. The slope of the curve in this figure is giving the Skempton coefficient $B$. Notice that for a precise evaluation of the Skempton coefficient, one has to account for the effect of the deformation of the drainage and pressure measurements systems. This point is discussed in details by Ghabezloo and Sulem [21] who showed that for the testing device used here, this correction is very small. We observe the linear increase of the pore pressure in the loading and reloading parts with the confining pressure increase. In the beginning of the unloading parts, the pore pressure decrease is retarded and the rate of change of pore pressure with the confining pressure is considerably lower than the one observed during the loading part. Below a certain stress level, about 15MPa, there is a considerable increase in the rate of decrease of the pore pressure and no excess pore pressure remains at the end of the unloading part. As it can be observed on Figure (13), the pore pressure variation in this part of the unloading curve is equal to the variation of the confining pressure. This indicates that the pore fluid has flowed between the sample and the rubber membrane and that consequently the slope of the curve in this part is not the Skempton coefficient. The same problem is observed in the beginning of the reloading parts. Because of this, the Skempton coefficients are evaluated on the central (linear) part of the reloading curves. Figure (14) presents the evaluated Skempton coefficient as a function of Terzaghi effective stress evaluated at each point and shows a very small reduction of this coefficient with the effective stress. This small variation can be neglected and the Skempton coefficient can be considered as constant, equal to the average of the measured values, $B = 0.4$.

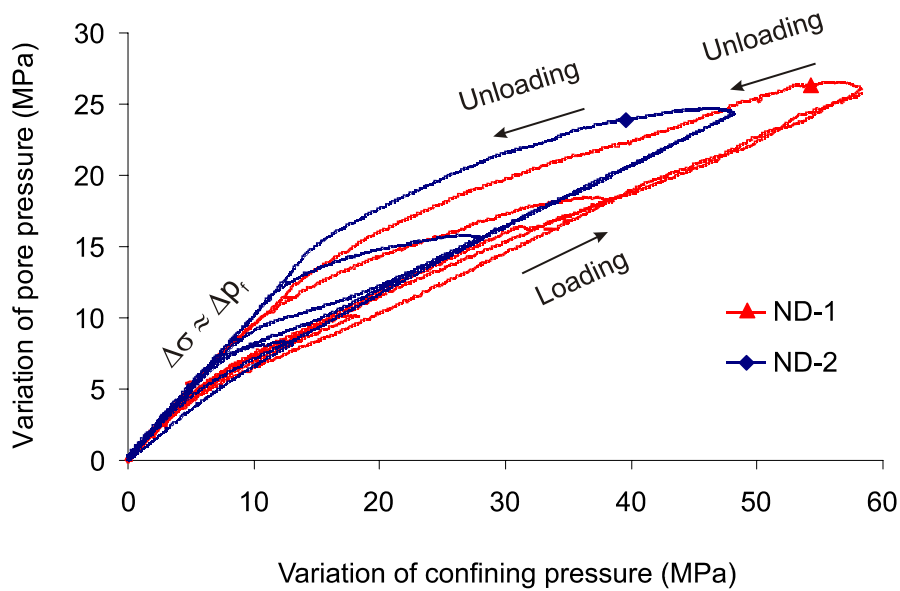

**Figure 13- Undrained isotropic compression tests, pore pressure evolution**

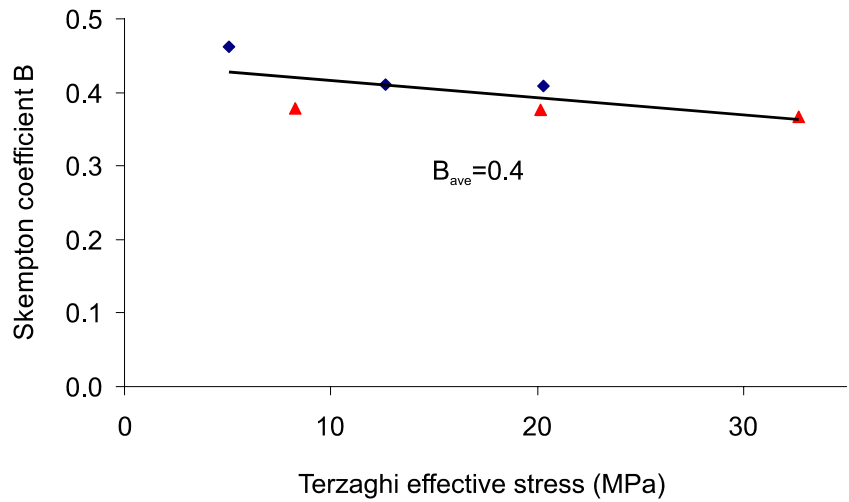

**Figure 14- Variation of the Skempton coefficient B with Terzaghi effective stress**

Assuming an ideal porous material for which $K_s = K_\phi$, the following relation is obtained for the Skempton coefficient by replacing equation (45) in equation (13):

$$\frac{1}{B} = 1 + \left(\frac{K_s}{K_f} - 1\right) \frac{\phi(1-\phi)^3}{1-(1-\phi)^3} \qquad (49)$$

Replacing $K_s = 21\,\text{GPa}$, $K_f = 2.2\,\text{GPa}$ and $\phi = 0.26$ in this relation, the Skempton coefficient is found equal to 0.40 which is equal to the average value obtained in the test results. Introducing equations (45) and (49) in equation (14), the undrained bulk modulus is found equal to 11.1GPa which is here again has a very good compatibility with the obtained experimental result, presented in equation (48). The good evaluation of the drained and undrained bulk moduli and also of the Skempton coefficient using simple homogenization relations (equations (45) and (49)) shows the efficiency of this method for evaluation of the poromechanical properties of heterogeneous materials, when used with proper parameters, and also the great importance of the unjacketed modulus $K_s$ in these evaluations.

## 3.7 *Summary of experimentally evaluated poroelastic parameters*

The evaluated poroelastic parameters of hardened cement paste are summarized below:

$$K_s = 21$$
$$K_d^{\text{sec}} = 8.69 - 0.087\sigma_d$$
$$K_u^{\text{sec}} = 11.25 - 0.099\sigma_d \quad (K_{s,d,u} \text{ in GPa and } \sigma_d \text{ in MPa}) \quad (50)$$
$$B = 0.4$$
$$\alpha^{\text{sec}} = 0.586 + 0.004\sigma_d$$

The graphical representation of the variations of the poroelastic parameters of hardened cement paste is presented in the Figure (15). Terzaghi effective stress in this figure is limited to 40MPa which is the maximum stress level achieved in the undrained tests.

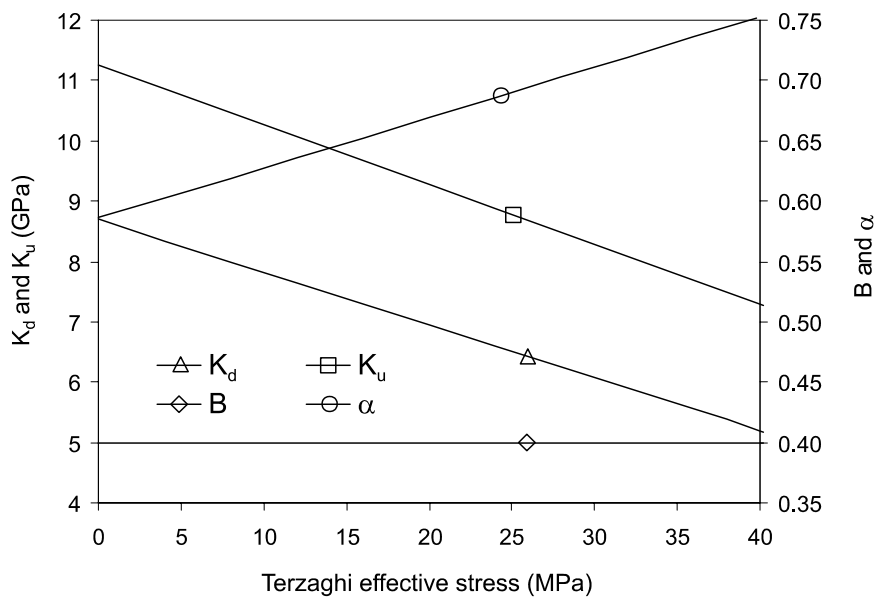

**Figure 15- Variation of poroelastic parameters of the hardened cement paste with Terzaghi effective stress**

# 4 Analysis of the overall compatibility of the experimental results

The Biot effective stress $\sigma'_b$ can be calculated using equations (43) and (46). The performed undrained compression tests can be analysed using Biot effective stress to calculate the drained bulk modulus and the results can be compared with the results of the drained tests to ensure the compatibility of the drained and the undrained isotropic compression tests. Figure (16) presents the $\varepsilon - \sigma'_b$ curves of the undrained tests together with the $\varepsilon - \sigma$ curves of the drained tests. In this figure we can see the consistency of the stress-strain curves of the drained and undrained tests. The drained bulk moduli evaluated in the undrained tests as described above, are presented in Figure (17) together with the drained bulk moduli evaluated in the drained compression tests. This graph shows the good compatibility of the bulk moduli calculated in the drained and the undrained tests, and also confirms the appropriate choice of the loading rate used for the drained tests.

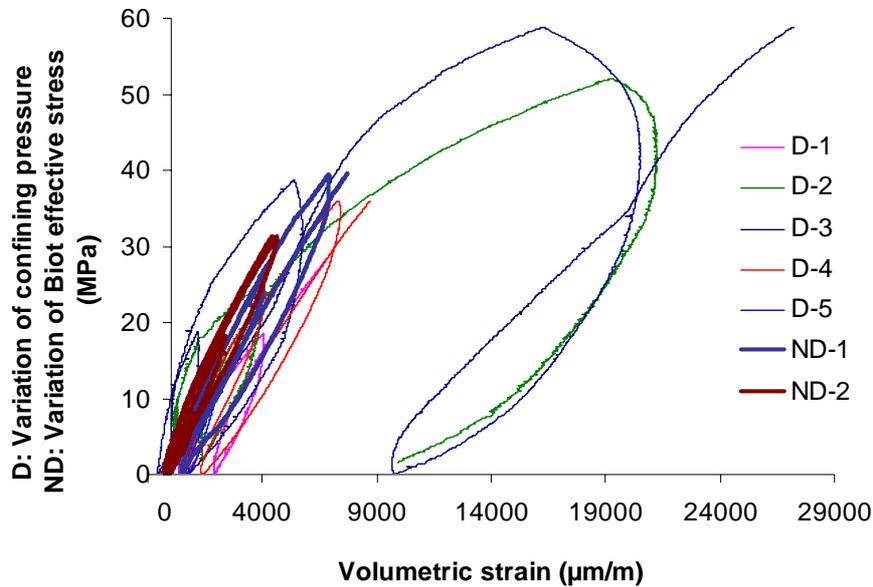

**Figure 16- Evolution of the volumetric strain with the confining pressure for drained tests (D) and with Biot effective stress for undrained tests (ND)**

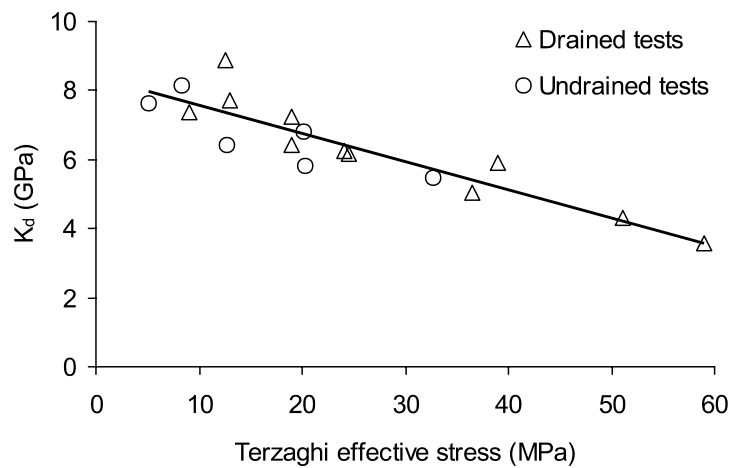

**Figure 17- Drained bulk modulus evaluated in drained and undrained tests as a function of Terzaghi effective stress**

In equation (46) the Biot effective stress coefficient is evaluated using the results of drained compression tests and unjacketed test. An other expression can be obtained using equations (14) and (25) to evaluate this coefficient using the results of drained and undrained compression tests:

$$\alpha = \frac{1 - K_d / K_u}{B} \quad (51)$$

By inserting equation (44) and (48), respectively for $K_d$ and $K_u$, and with $B = 0.4$, the Biot coefficient can be evaluated using equation (51). Comparing the Biot coefficients evaluated using equations (46) and (51), (see Figure (18)), we can observe the good compatibility of the experimental results obtained from the drained, undrained and unjacketed compression tests.

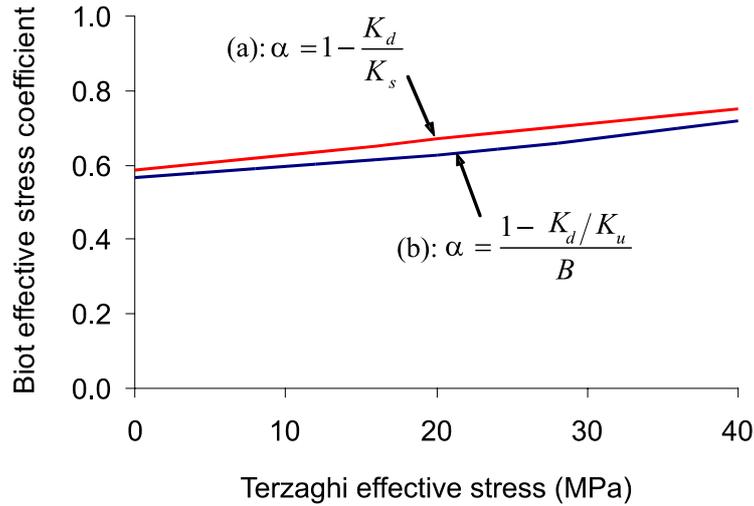

**Figure 18- Comparison of Biot effective stress coefficients evaluated from (a) drained and unjacketed compression tests and (b) drained and undrained compression tests**

Introducing the expression of the Biot coefficient presented in equation (25) in equation (51), the following expression is obtained for evaluation of the unjacketed modulus $K_s$ using the results of drained and undrained tests:

$$K_s = \frac{K_d K_u B}{K_d - K_u (1 - B)} \quad (52)$$

The differential pressure during the performed unjacketed test was maintained constant equal to 0.5MPa. Using this value in equations (44) and (48), the drained and undrained bulk moduli are respectively evaluated equal to 8.65GPa and 11.20GPa. The Skempton coefficient is constant, equal to 0.4. Inserting these values in equation (52), the unjacketed modulus is evaluated equal to 20.1GPa which is compatible with the measured value equal to 21GPa and shows again the good compatibility of the obtained experimental results.

Note that the above discussion on the compatibility of the various experimental results is not affected by the cement paste porosity.

# 5   Analysis of the results

Using the obtained experimental results an interesting analysis can be done on the cement paste porosity that should be used in poromechanical fromulations, as discussed in section 2.4. Replacing equations (25) and (30) in inequality $\kappa \leq \alpha$ of equation (31) the following inequality is obtained:

$$\frac{1}{K_\phi} \leq \frac{\alpha}{\phi K_s} \qquad (53)$$

Inserting inequality (53) in equation (16) the following inequalities are obtained for the cement paste porosity:

$$\phi \leq \begin{cases} K_f \left( \dfrac{\alpha}{K_s} + \dfrac{(1/K_d - 1/K_s)(1/K_u - 1/K_s)}{(1/K_d - 1/K_u)} \right) & \text{(a)} \\ K_f \left( \dfrac{\alpha}{K_s} + \dfrac{(1-B)(1/K_d - 1/K_s)}{B} \right) & \text{(b)} \\ K_f \left( \dfrac{\alpha}{K_s} + \dfrac{1/K_u - 1/K_s}{B} \right) & \text{(c)} \\ K_f \left( \dfrac{\alpha}{K_s} + \dfrac{(1-B)(1/K_d - 1/K_u)}{B^2} \right) & \text{(d)} \end{cases} \qquad (54)$$

Replacing the values of poroelastic parameters, $K_d = 8.69\text{GPa}$, $K_u = 11.25\text{GPa}$, $K_s = 21\text{GPa}$, $B = 0.4$, $\alpha = 0.586$ and $K_f = 2.2\text{GPa}$ in equation (54) we obtain $\phi \leq 0.29$ for equations (a) and (c) and $\phi \leq 0.28$ for equations (b) and (d). The obtained porosity upper limits are lower than the total porosity of the studied cement paste, evaluated by oven drying equal to 0.35. This confirms the discussion presented in section 2.4 and shows that the cement paste porosity that should be used in poromechanical formulations is smaller than the total porosity, which contains some quantity of interlayer water.

The obtained poroelastic parameters can be used in equation (16) to evaluate the modulus $K_\phi$. This evaluation is done for a zero differential pressure. Inserting the above mentioned values of poroelastic parameters with $\phi = 0.26$ in equation (16), the parameter $K_\phi$ is evaluated equal to 21.9GPa for equation (a), 15.3GPa for equation (b), 17.3GPa for equation (c) and 13.0GPa for equation (d). We can see that despite of the good compatibility of the experimentally evaluated parameters, as shown in section 4, the indirect evaluation of the modulus $K_\phi$ is very difficult and the obtained values differ significantly. However, we consider the average of the obtained values for the modulus $K_\phi$, equal to 16.9GPa, which is smaller than the experimentally evaluated unjacketed modulus $K_s$, equal to 21GPa. It should be noted that for an ideal porous material which is homogeneous and isotropic in the micro-scale, the parameters $K_\phi$ and $K_s$ are equal.

Assuming that the modulus $K_\phi$ is constant, we can evaluate the effective stress coefficients $\beta$, $\chi$ and $\kappa$ defined respectively in equations (26), (28) and (30) as functions of Terzaghi effective stress. Figure (19) presents the variations of these effective stress coefficients, and also of $\alpha$ and $\theta$, with Terzaghi effective stress and we can observe on this graph that the inequality (31) is verified for the evaluated parameters.

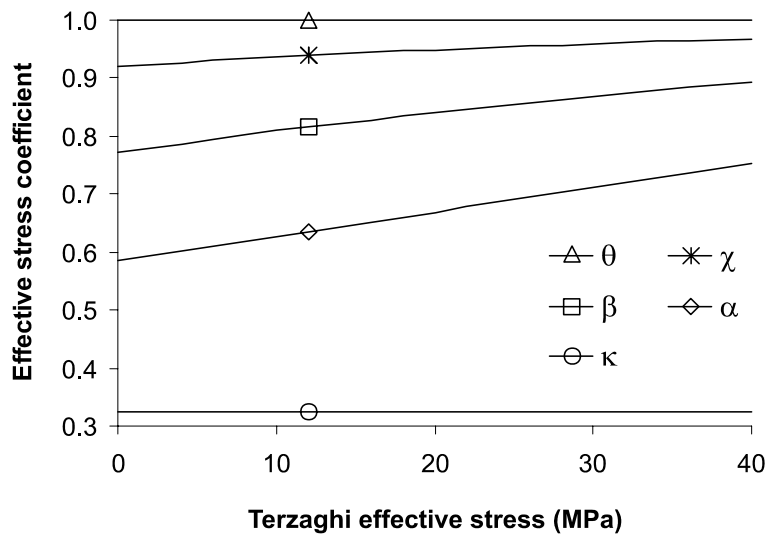

**Figure 19- Effective stress coefficients corresponding to different physical properties as a function of Terzaghi effective stress ($\theta$ for drained bulk modulus, $\chi$ for porosity, $\beta$ for pore volume, $\alpha$ for total volume and $\kappa$ for solid volume)**

As mentioned earlier, in this study we have assumed that the undrained parameters, $K_u$ and $B$, vary with Terzaghi effective stress. Using the obtained average value of $K_\phi$ it is now possible to study the influence of this assumption on the above analysis. This is done by studying the effect of deviation of $K_\phi$ from $K_s$ on the evaluation of Skempton coefficient. Inserting the poroelastic parameters presented above in equation (13), the Skempton coefficient is evaluated equal to 0.396 which has a very good compatibility with the measured value, equal to 0.4. Replacing $K_\phi$ with $K_s$, the evaluated value is reduced to 0.389, i.e. a change of less than 2% in the evaluated Skempton coefficient. Consequently, we can conclude that the assumption that the undrained parameters vary with Terzaghi effective stress does not influence significantly the obtained results.

# 6 Microscopic observations

Microscopic observations are performed on the hardened cement paste samples before and after the isotropic compression tests in order to study possible changes in the microstructure of the cement paste that could corroborate the non-linear stress-strain behaviour and degradation mechanism of the elastic moduli observed in isotropic compression tests.

## 6.1 Observation equipment and conditions

The observations are performed using an environmental scanning electron microscope (ESEM) which is equipped by two types of detectors: A detector of secondary electrons (SE) and a detector of backscattered electrons (BSE). The SE detector results in images which emphasize the topographical contrasts on the observed sample and have a three-dimensional appearance. In these images, steep surfaces and edges tend to be brighter than flat surfaces. The BSE detector can detect the contrast between areas with different chemical compositions, especially when the average atomic number of the various regions is different. The brightness of the images taken by this detector tends to increase with the atomic number. In order to keep the natural contrast of the samples, the observation conditions are adapted to work in low vacuum condition.

## 6.2 Sample preparation

The operations of sample preparation and microscopic observation are performed shortly after each mechanical loading test in order to avoid the effects of carbonation on the sample. Working in the low vacuum condition, it is possible to observe the samples without drying them. Two types of samples were used for microscopic observations: Fractured samples and polished samples. Fractured samples are prepared by breaking the sample with a single hammer blow on a burin placed on the sample. The SE detector is used to take the images of these samples. Preparation of polished samples was performed following the normal procedure of preparation of flat-polished samples for SEM observations, including cutting, grinding and polishing steps. The details of such a procedure are presented by Stutzman and Clifton [51] and Kjellsen et al. [52]. The images of these samples are taken using the BSE detector.

## *6.3 Observation results*

Figure (20) shows an image of the microstructure of the hardened cement paste taken on a polished intact sample with 200x magnification. As explained by Diamond [53], we can observe the unhydrated cement particles as bright entities, hydration shells surrounding the unhydrated cores and fully hydrated cement grains as smooth-textured uniformly grey areas. The hydration shells and fully hydrated grains, formed in originally grain-filled space, are called 'inner products' and appear non-porous at this magnification. These inner products are separated from each other by a groundmass or 'outer product' deposited in the originally water-filled space which is irregularly textured and has a much less homogeneous appearance. This image clearly shows the very heterogeneous microstructure of the hardened cement paste.

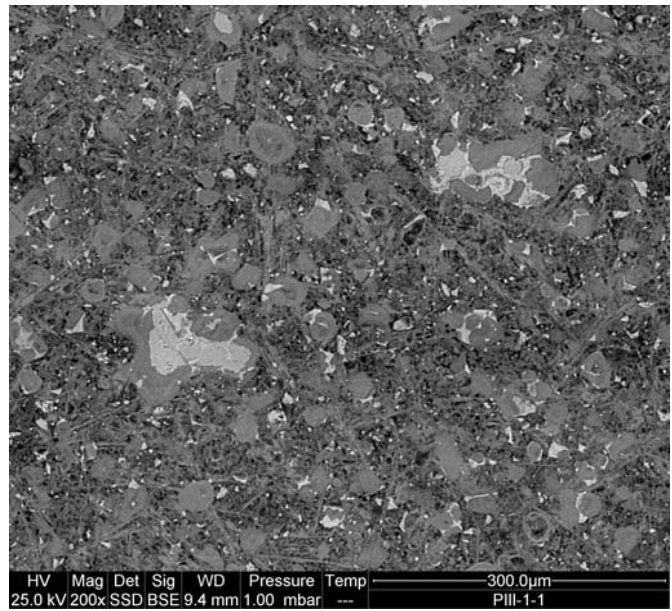

**Figure 20- Microphotograph of the heterogeneous microstructure of the hardened cement paste, image taken on an intact, polished sample**

Figure (21) shows an image taken after a drained isotropic compression test on a fractured sample with 400x magnification and we can see clearly the formation of a microcrack in the cement paste. Several images taken on different samples after the drained compression tests did not show any preferential direction for the microcracks. The length of the observed microcracks is generally between 100 and 200 µm. To ensure that these microcracks are not produced during the preparation of the sample, the same preparation method by fracturing is applied on two intact samples. The microscopic observation of these samples did not show any microcracks. Figure (22) shows another image taken after a drained compression test on a polished sample with 50x magnification. We can see on this image, an other type of inhomogeneity of the hardened cement paste in the form of areas with different grain densities and porosities, exhibited at a greater scale as compared to the microstructural inhomogeneity observed in Figure (20). This type of inhomogeneity has been also observed by Diamond [53]. On this image we can also observe the rupture of an air void situated in the more porous area. The traces of the polishing operation performed during the sample preparation can be observed as straight lines in the left part of the image presented in Figure (22). These traces are parallel and cannot be confounded with eventual microcracks generated during the mechanical loading test. Another example of the rupture of an air void is presented in Figure (23), taken on a polished sample after a drained compression test. Microscopic observation of the samples after the undrained compression tests shows the same type of microcracking of the samples under isotropic loading. An example, taken on a fractured surface with 1200x magnification, is presented in Figure (24) and shows a microcrack with an opening of a few microns. It should be mentioned that this opening is observed after the test on the unloaded sample.

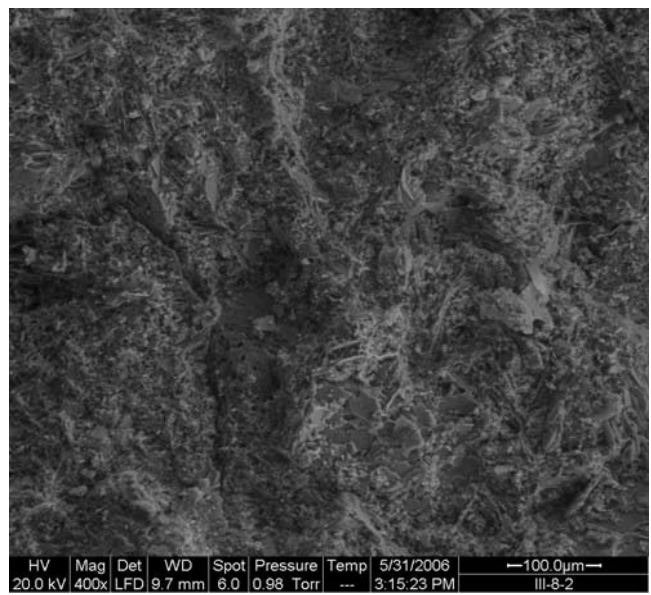

**Figure 21- Microcracking of the sample, image taken on a fractured sample after a drained isotropic compression test**

These images show that the two different observation methods used in this study, backscattered electron SEM imaging on the polished samples and secondary electron SEM imaging on the fractured samples, are two complementary methods for the observation of the *Ghabezloo et al.: Poromechanical behaviour of hardened cement paste under isotropic loading*microcracked samples. The images taken on the polished samples show clearly the heterogeneity of the microstructure of the hardened cement paste and the details of the rupture of the microstructure near the air voids. But the observation and detection of the microcracks on these images is more difficult than on the images taken on the fractured samples. These latter provide a three-dimensional and very clear view of the existing microcracks.

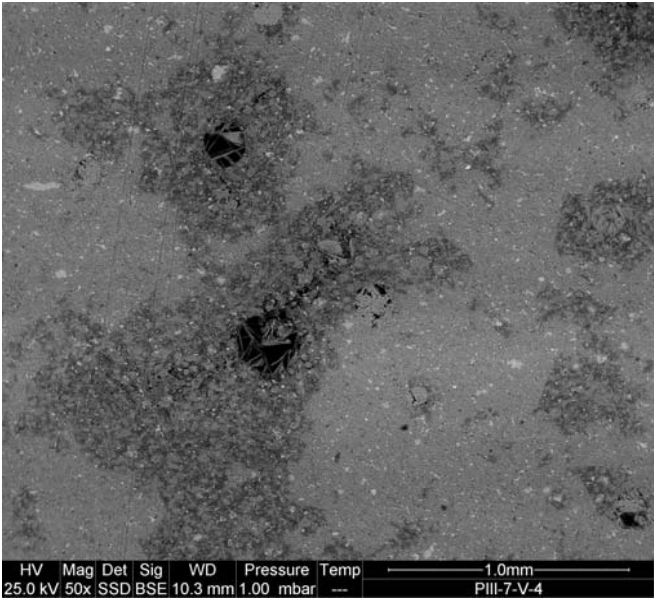

**Figure 22- Microphotograph of the heterogeneous microstructure of the hardened cement paste and the rupture of the air voids, image taken on a polished sample after a drained isotropic compression test**

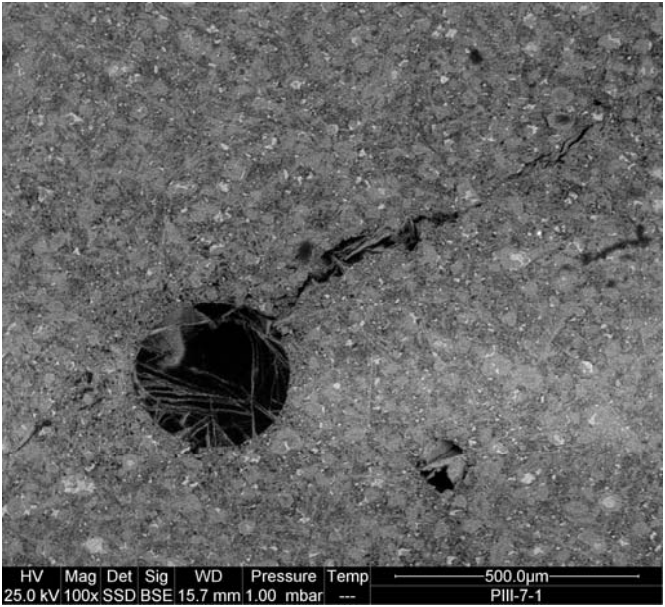

**Figure 23- Rupture of an air void, image taken on a polished sample after a drained isotropic compression test**

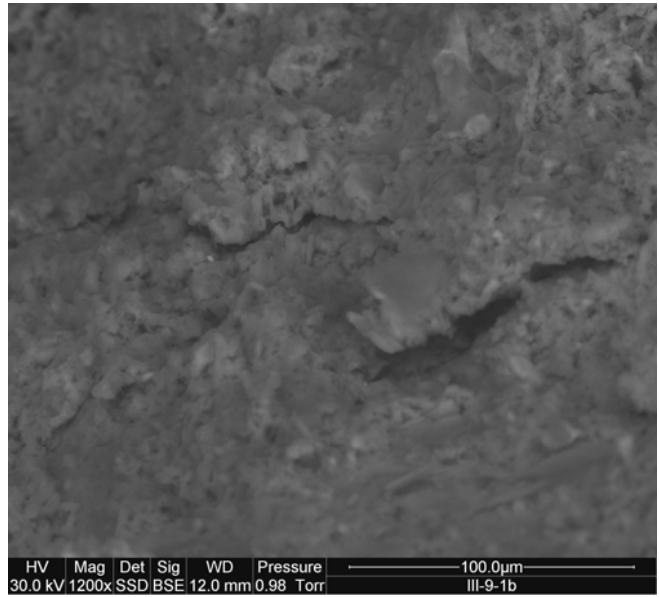

**Figure 24- Microcracking of the sample, image taken on a fractured sample after an undrained isotropic compression test**

# 7 Mechanism of degradation under isotropic loading

The microscopic observations presented in the preceding section showed the presence of two types of heterogeneities in the microstructure of the hardened cement paste. The first type of inhomogeneity (Figure (20)), exhibited at a smaller scale, is due to the presence of different phases, including unhydrated grains, low density and high density C-S-H, CH crystals and the macro-porosity in the microstructure of the cement paste. These phases have different mechanical properties which are evaluated by different experimental methods, as in particular the nanoindentation method, by Monterio and Chang [2], Acker [3], Le Bellego [4], Velez et al. [5], Constantinides and Ulm [6]. Ulm et al. [1] have presented a table which summarizes the results obtained in these studies which can be used to evaluate the bulk modulus of the different phases of the microstructure of the hardened cement paste. The bulk modulus is evaluated equal to 14GPa and 19GPa respectively for low density and high density C-S-H, 32GPa for CH crystals and between 104GPa and 121GPa for unhydrated grains. The unjacketed modulus of the hardened cement paste, evaluated equal to 21GPa in our experiments, is some weighted average of the bulk moduli of different phases of the microstructure. The presence in the microstructure of the hardened cement paste of various materials with bulk moduli which differ by one order of magnitude may lead to incompatible deformations at the interface between the various materials which can induce microcracking and degradation of the mechanical properties even under isotropic loading.

The second type of inhomogeneity (Figure (22)) is presented at a greater scale as compared to the first type in the form of areas with different porosities and therefore with different average mechanical properties. The difference between the average elastic moduli of two adjacent areas in the microstructure can also cause microcracking and degradation of the material under isotropic loading. This can be seen in the images showing the rupture of the air voids presented in the Figure (22).

# 8 Conclusions

The poromechanical behaviour of hardened cement paste under isotropic loading is studied on the basis of an experimental testing program of drained, undrained and unjacketed compression tests. The macroscopic behaviour of the material is described in the framework of the mechanics of porous media. The poroelastic parameters of the material are determined and the effect of stress and pore pressure on them is evaluated. The unjacketed modulus of hardened cement paste is evaluated equal to 21GPa and the tests results show that it does not vary with the applied pressures. Secant drained and undrained bulk moduli and the Skempton coefficient of the material are evaluated in unloading-reloading cycles at different stress levels during the isotropic drained and undrained compression tests. Drained compression tests are performed with different pore pressures and show that the drained bulk modulus is a function of Terzaghi effective stress. Test results show the reduction of drained and undrained bulk moduli with Terzaghi effective stress increase. The microscopic observation of the samples showed that this degradation phenomenon is caused by the microcracking of the material under isotropic loading. The microcracking can be caused by the heterogeneity of the microstructure of the hardened cement paste which manifests itself at different scales. The Skempton coefficient is found to be constant, equal to 0.4. Appropriate effective stress laws which control the evolution of total volume, pore volume, solid volume and porosity of the material are discussed. The Biot effective stress coefficient increases linearly with Terzaghi effective stress. Analysis of the experimental results showed that the cement paste total porosity, measured by oven-drying at 105°C, can not be used in poromechanical formulations. The drained and undrained bulk moduli, the Skempton coefficient and the Biot effective stress coefficient are well reproduced with simple homogenization equations using the evaluated unjacketed modulus and the porosity. Finally, the good compatibility and consistency of the obtained poromechanical parameters demonstrates that the behaviour of the hardened cement paste can be indeed described within the framework of the theory of porous media.

# 9 Acknowledgments

The authors gratefully acknowledge TOTAL for supporting this research. They wish also to thank Xavier Boulay for performing mercury porosimetry experiments.